\documentclass[twocolumn]{article}
\usepackage{exscale}                  
\usepackage[intlimits]{amsmath}       
\usepackage{amsfonts}
\usepackage{amssymb,amscd}
\usepackage[dvips]{epsfig}                   
\usepackage{array}
\usepackage{wrapfig}
\usepackage[usenames]{color}
\usepackage{graphicx}
\usepackage{bm}
\def\ksl{k\hspace{-0.2cm}/}

\usepackage{graphics}
\usepackage{graphicx}
\usepackage{dcolumn}

\newcommand{\ar}{\arrowvert}
\newcommand{\ra}{\rangle}
\newcommand{\la}{\langle}
\newcommand{\da}{\dagger}

\newcommand{\cd}{\! \cdot \!}
\newcommand{\be}{\begin{equation}}
\newcommand{\ee}{\end{equation}}
\newcommand{\ba}{\begin{eqnarray}}
\newcommand{\ea}{\end{eqnarray}}
\newcommand{\bs}{\boldsymbol} 
\begin{document}

\title{Parity doubling in the high baryon spectrum:\\
near-degenerate three-quark quartets}
\author{Felipe J. Llanes-Estrada$ ^1$\footnote{Part of a 
talk delivered at the miniworkshop Bled-2008}
,Pedro Bicudo$ ^2$, Marco 
Cardoso$ ^2$, Tim Van Cauteren$ ^3$\\
$ ^1$Dept. F\'{\i}sica Teorica I, Univ. Complutense, Madrid
28040, Spain\\
$ ^2$Instituto Superior Tecnico,1049-001 Lisboa, Portugal \\
$ ^3$Dept. Subatomic and Radiation Physics, Ghent University, Belgium}

\date{October 15th, 2008}
\maketitle

\begin{abstract}
We report on the first calculation of excited baryons with a 
chirally symmetric Hamiltonian, modeled after Coulomb gauge 
QCD (or upgraded from the Cornell meson potential model to a 
field theory in all of Fock-space) showing the
insensitivity to chiral symmetry breaking.
As has recently been understood, this  leads to doubling 
between two hadrons of equal spin and opposite parity. 
As a novelty we show that three-quark $\Delta$ states group 
into quartets with two states of each parity, all four states having equal angular momentum $J$. 
Diagonalizing the chiral charge expressed in terms of quarks 
we show that the quartet is slightly split into two parity doublets 
by the tensor force, all splittings decreasing to zero high in 
the spectrum. \\
Our specific calculation is for the family of maximum-spin 
excitations of the Delta baryon.
We provide a model estimate of the experimental accuracy
needed to establish Chiral Symmetry Restoration in the high
spectrum. 
 We suggest that a measurement of masses of high-partial wave 
$\Delta$ resonances with an accuracy
of 50 MeV should be sufficient to unambiguously establish the 
approximate degeneracy, and test the concept of running quark 
mass in the infrared. 
\end{abstract}

\vspace{0.3cm}
\newpage
The idea of chiral symmetry restoration has been
around for a while, for example parity doubling 
was examined for the proton in the context of the 
linear sigma model in  \cite{Detar:1988kn}. By 
current ideas we believe that this restoration 
should occur for higher excitations.
Glozman and collaborators 
\cite{Glozman:1999tk,Glozman:2002jf,Wagenbrunn:2006cs,Cohen:2005am,Cohen:2001gb,Cohen:2002st,Glozman:2005tq}
(see also \cite{Jido:1999hd})
have theoretically examined ($q\bar{q}$) 
mesons, and also  shown marginal empirical 
evidence for chiral symmetry restoration in  both meson and hadron 
spectra,  that rekindles interest
on intermediate energy resonances.  Chiral symmetry 
restoration, or more precisely,
Spontaneous Chiral Symmetry Breaking Insensitivity high in the spectrum,
is established as a strong prediction of the symmetry breaking pattern of QCD, 
and such prediction in an energy region where little else can be stated,
needs to be confirmed or refuted by experiment.
\\
The baryon spectrum is a more difficult theoretical problem given the 
minimum three-body wavefunction (as opposed to only quark-antiquark for
mesons) and in this paper we provide the necessary theoretical 
background to understand parity doubling, in agreement with a prior 
study by Nefediev, Ribeiro and Szczepaniak \cite{Nefediev:2008dv}, and
give the first model estimate of what the experimental target-precision should be. 
This should help quantify what  ``high enough'' in the spectrum means,
to assist experimental planning. 
\\

\begin{figure}[h]
\centerline{\epsfig{file=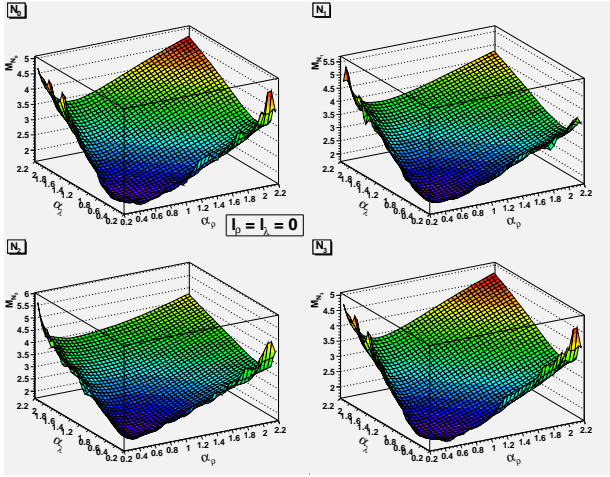,width=3.2in}}
\caption{Variational minimum-energy search 
$E(\alpha_\rho,\alpha_\lambda)$ with a two-parameter family of
functions. Best results are obtained when the (chiral-limit) pion 
wavefunction is rescaled and used to build the Jacobi-radial 
part of the $\Delta$ wavefunctions,
$\sin\phi(k_\rho/\alpha_\rho) \sin\phi(k_\lambda/\alpha_\lambda)$. For 
maximum spin $\Delta$ states, $J=3/2+l_\rho$ the angular wavefunction before symmetrization
is $Y_l^{m_l}(\hat{k}_\rho)$ (we set $l_\lambda=0$ consistent with the
variational approximation, but numerically symmetrize the spin-space
wavefunction, which reintroduces it through exchanged-quarks).
\label{fig:varsearch}}
\end{figure}
\begin{figure}[h]
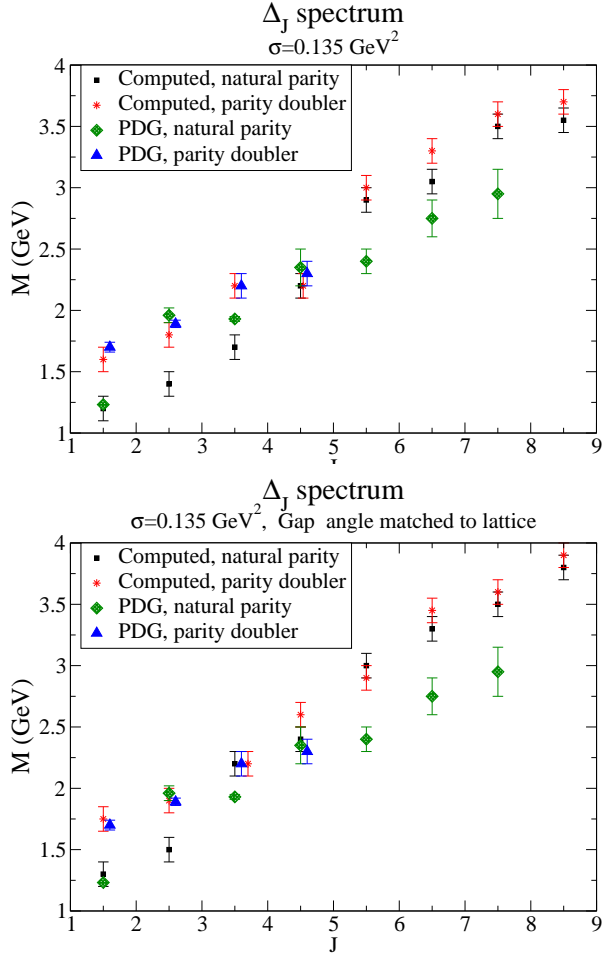

\centerline{\epsfig{file=lowstring.eps,width=3.1in}}
\centerline{\epsfig{file=lowstring2.eps,width=3.1in}}
\caption{Parity doubling in the spin-excited $\Delta$ spectrum.
Top: with infrared quark mass as calculated in the model 
(probably too low). Bottom: quark mass rescaled to fit Landau-
gauge lattice data. The model clearly displays parity 
doubling.
The experimental situation is still unclear, 
the degeneracy can be claimed for the $9/2$ states alone, and the chiral
partners higher in the spectrum are not experimentally known. 
\label{fig:doubling}}
\end{figure}
\begin{figure}[h]
\centerline{\epsfig{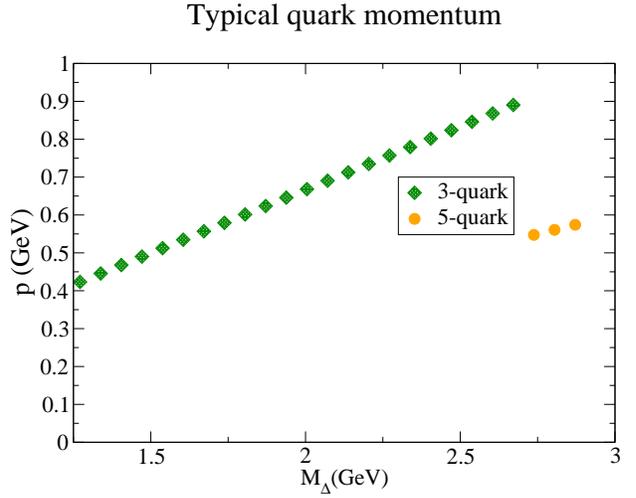}}
\caption{
The typical momentum of a quark in a three-quark state is (by kinetic 
energy considerations alone, with a running mass-gap) $\ar k \ra\propto
M_{\Delta_J}$. Plotted is the typical momentum in a three quarks and 
five quark wavefunction. At the jump the phase space for five-quark
states is larger, so it is more likely that a baryon of that mass
is in a five-quark configuration, and the typical momentum is therefore 
smaller. Hence chiral symmetry restoration has to be somewhat slower
than three-quark models would indicate. \label{fig:typicalp}
}
\end{figure}
\begin{figure}[h]
\centerline{
\epsfig{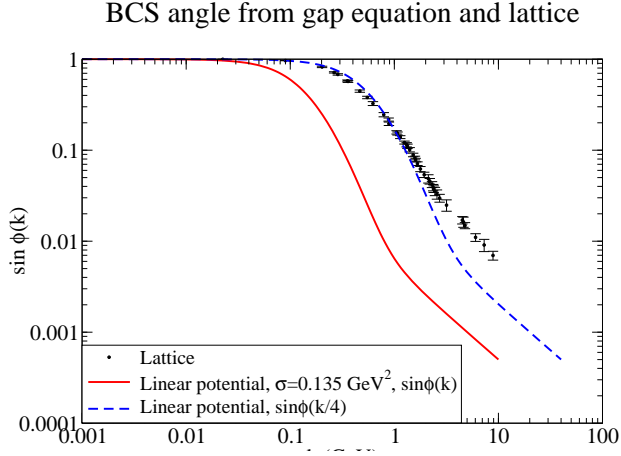}}
\centerline{\epsfig{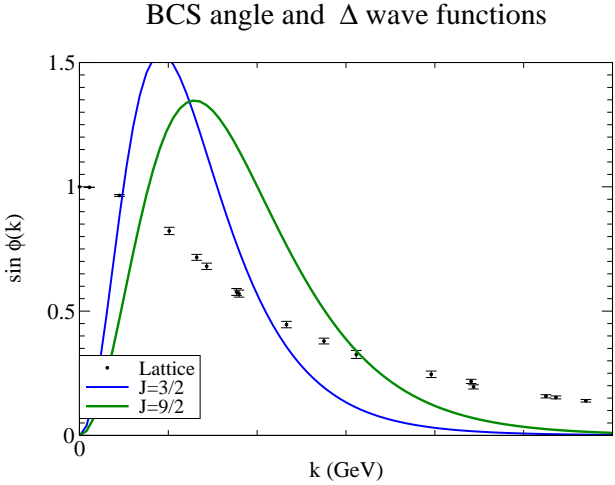}}
\caption{The sine of the gap angle $M(k)/\sqrt{(M(k)^2+k^2)}$ has
  limited support if the chiral-symmetry breaking quark mass remains
  of order $\Lambda_{QCD}$ or less. Top: we show the
  running mass from a model computation for a linear potential with
  string tension $\sigma=0.135\ GeV^2$, and its rescaling to match
  Landau-gauge data\cite{Bowman:2006zk,Parappilly:2005ei} (no
  Coulomb-gauge lattice data for the quark mass is known to us).
 Bottom: Quark-momentum distributions for
  $\Delta_{3/2}$ and $\Delta_{9/2}$ with simple variational
  wavefunctions.  The quark-momentum distribution for higher hadron
  resonances has smaller overlap with this gap angle, and therefore
  the quarks in those hadrons behave effectively as if they were
  massless. Hence they become insensitive to the gap angle, and chiral
  symmetry is restored in Wigner-Weyl mode with degenerate multiplets.
\label{fig:restoration}}
\end{figure}

We customarily employ a truncation of Coulomb-gauge QCD by
ignoring the Faddeev-Popov operator and substituting the 
Coulomb  kernel by its vacuum expectation value, that takes 
the usual linear plus Coulomb form. 
This can be seen as a field theory upgrade of the Cornell 
potential model.  
The Hamiltonian reads
\begin{eqnarray}  \nonumber
H = - g_s \int d{\bf x} \Psi^\da (x) {\bf \alpha}  \cd {\bf A}(x) 
\Psi (x) \\ \nonumber
 + Tr \int d{\bf x} ( {\bf E}  \cd  {\bf E} +
{\bf B}  \cd {\bf B} )
\\ \nonumber
+
\int d {\bf x} \Psi ^{\dagger} _q
({\bf x}) (-i {\bf \alpha} \cd \nabla + \beta m_q )
\Psi_q({\bf x}) \\
+ \frac{1}{2} \int d {\bf x} d {\bf y}
\rho^a({\bf x})V_L(\arrowvert {\bf x} -
{\bf y} \arrowvert) \rho^a({\bf y})  
\end{eqnarray}
with a strong kernel containing a linear potential $V_L$, with
string tension $\sigma=0.135\ GeV^2$, coupled
to the color charge density 
$
\rho^a({\bs x}) = \Psi^\dagger({\bs x}) T^a\Psi({\bs x}) +f^{abc}{\bf
A}^b({\bs x})\cdot{\bf \Pi}^c({\bs x}) \ .
$
In our past work we have solved the BCS gap equation to 
spontaneously break chiral symmetry.
This model has the same chiral structure of QCD, satisfying 
the Gell-Mann-Oakes-Renner relation, the low-energy theorems 
for pion scattering\cite{Bicudo:2001jq}
and allowing computations of static pion-nucleon observables\cite{Bicudo:2001cg}.
We have employed it in studies of 
gluodynamics\cite{Bicudo:2006sd} shown at this workshop that 
agree with lattice gauge theory and are of qualitative 
phenomenological interest. In
any case, these play a minor role in the topic of this article, as
the decreasing of the splittings is dominated by chiral symmetry breaking
alone.
For a reduced baryon sector application we are going to 
perform two  more simplifications. We employ only the $V_L$ 
linear potential, and
neglect all magnetic interactions. This makes the $\Delta$-nucleon mass
splitting too small, but does not affect the $\Delta$ spectrum much. 

We truncate the Fock space variationally, as customary, to the $\ar 
qqq\ra$ minimum wavefunction. Since radial excitations of this system
compete with multiquark excitations, we concentrate instead  on
maximum angular-momentum excitations $J=3/2+l$. Chiral forces are too weak
to compensate large centrifugal forces and can hardly maintain $l=3$ or $l=4$,
so one hopes to reduce the molecular component by studying the ground state
in each $J$-channel, so that the $\ar qqq\ra$ correlation remains important
high in the spectrum. 

 As a rule of thumb, one needs to keep in the 
 Fock-space expansion $\ar qqq\ra + \ar qqqq\bar{q}\ra
+\ar qqqg\ra+\dots$.
as many states as will be competitive by phase space considerations, considering
the quark and gluon dynamical mass gaps established by lattice and Dyson-Schwinger
studies. When pentaquark correlations are more abundant than 
three-quark correlations (see figure \ref{fig:typicalp}) the 
typical quark momentum will be lower than extrapolated from 
the ground-state baryons, so that chiral symmetry restoration 
will not be quite so fast.

This puts pentaquark correlations above 2$GeV$, with the exception
of possible meson-baryon resonances (as the Goldstone bosons avoid the mass-gap).
In any case it seems well established that three-quark correlations play
an important role in baryon-phenomenology, so it is worth examining the
effect of a chiral transformation on a three-quark variational wavefunction
$\ar N \ra =F_{ijk} B^\da_i B^\da_j B^\da_k \ar 0 \ra$.

We proceed variationally and employ several types of wavefunctions, rational
and Gaussian, but the lowest energy (binding the model's $J$-ground state from above by the Rayleigh-Ritz principle) is obtained by employing the chiral limit pion-wavefunction rescaled with two variational parameters
in terms of Jacobi coordinates, $\sin \phi (k_\rho/\alpha_\rho)
\sin \phi (k_\lambda/\alpha_\lambda) Y_l^{m_l}(\hat{k}_\rho)$. We have
found the angular excitation in $\lambda$ to be slightly higher in energy
and neglect the correlation. Part of it though reenters the calculation
upon (anti)symmetrizing the wavefunction, since quark exchange mixes the
$\rho$ and $\lambda$ variables.
A typical variational search is represented in figure
  \ref{fig:varsearch}.
Table \ref{tablita} presents the intradoublet splittings.
The interdoublet splittings, as well as improved precision on our three-body variational Montecarlo method, will be given in an upcoming publication.  
As can be seen from the table, the model doublet splittings drop with
the orbital angular momentum.
This is easy to understand from the structure of the model Hamiltonian.
The kernel for baryons is proportional to
\ba F^*_{s_1s_2s_3} \left({\bf k}_1,{\bf k}_2 \right)
U^\da_{k_1s_1}U_{k_1+q \lambda_1} \
U^\da_{k_2s_2}U_{k_2-q \lambda_2} \\ \nonumber \times F_{\lambda_1\lambda_2s_3} \left({\bf
  k}_1+{\bf q},{\bf k}_2-{\bf q} \right)\ea 
that, upon becoming insensitive to the gap angle, $\sin\phi(k>>\Lambda_{QCD})\to 0$,
turns into
\ba \label{eqchiral}
F^*_{s_1s_2s_3}
\left(\delta_{s_1\lambda_1} +(\sigma\cd \hat{k}_1 \sigma\cd 
\widehat{k_1+q})_{s_1\lambda_1}\right) \cd 
\\ \nonumber 
\left(\delta_{s_2\lambda_2} +(\sigma\cd \hat{k}_2\sigma\cd 
\widehat{k_2-q})_{s_2\lambda_2}\right) 
 F_{\lambda_1\lambda_2s_3}\ .
\ea
If instead of $F^*_{s_1s_2s_3}$ one substitutes its chiral partner
$F^*_{s_1's_2s_3}(\sigma\cd\hat{k}_1)_{s_1's_1}$ (and the same for the ket),
the two states are seen to be degenerate. Also apparent in
 Eq.(\ref{eqchiral}) is the role of
the tensor force in  enforcing chiral cancellations.

  Finally, the first 
computation of the parity doubling for baryons is presented in figure \ref{fig:doubling}.
\\

\begin{table}\caption{Experimental and computed doublet
 splittings.
 The entire quartet degenerates high in the spectrum, with the $+-$
 parity doubling proceeding faster due to insensitivity to $\chi$SB 
and the interdoublet splitting decreasing slower, as they are
due  to the tensor force and dynamical. We give a preliminary calculation of the intradoublet splitting (parity degeneracy). From the decreasing  theory splittings we deduce that an experimental measurement of the parity splitting $M_+ -M_-$ to an accuracy of 100, or better 50 MeV, should suffice to see the effect.
Note that our excited splittings become compatible with zero within errors in the Montecarlo  9-d integral.
\label{tablita}\vspace{0.1cm}}
\begin{tabular}{c|c|c}\hline
J & Exp.         & Theory  \\ 
  &     $M_+ -M_-$  &intradoublet  \\
\hline
3/2 & 470(40) &  450(100)  \\
5/2 & 70(90)  &  400(100)  \\
7/2 & 270(120)&   50(100) \\
9/2 & 50(250) &  200(100)  \\
11/2 & -      &  100(100)  \\
13/2 & -      &  100(100)  \\
\end{tabular}
\end{table}
%

Let us now show that there are indeed two closely separated baryon 
doublets, slightly split by tensor forces.
We find convenient to employ the gap angle instead of the quark mass
$$
\sin \phi (k) \equiv \frac{M(k)}{\sqrt{M(k)^2+k^2}}
$$
and the Dirac spinors can be easily parametrized as
\ba
U_{\kappa\lambda} = \frac{1}{\sqrt{2}} \left[ \begin{array}{c}
    \sqrt{1+\sin{\phi_\kappa}}\chi_\lambda \\ \sqrt{1-\sin{\phi_\kappa}}
    \vec{\sigma}\cdot\hat{\kappa} \chi_\lambda\end{array} \right]  \\
V_{-\kappa\lambda} = \frac{1}{\sqrt{2}} \left[ \begin{array}{c}
    -\sqrt{1-\sin{\phi_\kappa}} \vec{\sigma}\cdot\hat{\kappa}i\sigma_2
\chi_\lambda  \\
    \sqrt{1+\sin{\phi_\kappa}}i\sigma_2  \chi_\lambda  \end{array}
\right]\; .
\ea
Substituting these spinors, and in terms of Bogoliubov-rotated quark and
antiquark normal modes $B$, $D$, the chiral charge takes the form
\begin{eqnarray} \label{chiralcharge}
Q^5_a  = \int \frac{d^3k}{(2\pi)^3} \sum_{\lambda \lambda ' f f'c} 
\left(  \frac{\tau^a}{2} \right)_{ff'}\\ \nonumber
\left(\cos \phi(k) \right. \\  \nonumber
({\bf \sigma}\cd{\bf \hat{k}})_{\lambda \lambda'}
 \left( B^\da_{k\lambda f c} B_{k\lambda'f'c} 
 + D^\da_{-k\lambda f c} D_{-k\lambda'f'c}
 \right) + \\ \nonumber  \sin \phi(k) \\ \nonumber \left.
(i\sigma_2)_{\lambda \lambda'} \
\left( B^{\da}_{k\lambda f c} D^\da_{-k\lambda'f'c}+
B_{k\lambda f c} D_{-k\lambda'f'c}
\right) \right)  \ .
\end{eqnarray}
In the presence of Spontaneous Chiral Symmetry Breaking, $\sin \phi(k)\not=0$,
and the two terms in the second line are responsible for the non-linear
realization of chiral symmetry in the spectrum.
One can see this by applying the chiral charge on a hadron state to collect
 the same hadron state plus a pion. As in Jaffe,
Pirjol and Scardiccio \cite{Jaffe:2005sq}, 
\be  \label{nonlineartrans}
[Q_5^a,N^\pm_i]= v_0(\pi^2) \epsilon_{abc}\pi^c \Theta_{ij}^b N^\pm_j \ .
\ee
(Here, $i$ and $j$ are the chiral multiplet 
indices). \\  Eq. (\ref{nonlineartrans} )
is easy to derive because the $i\sigma_2$ matrix couples the quark-antiquark
pair to pseudoscalar quantum numbers, so the terms in the second line of
eq.(\ref{chiralcharge}) provide an interpolating field for the pion. In fact,
if the vacuum is variationally chosen as the BCS ground state $\ar \Omega\ra$
with $B\ar \Omega\ra=0$,
$D\ar \Omega \ra=0$, $\sin \phi(k)$ then provides precisely the RPA pion wavefunction
in the chiral limit, and the terms with $\sin \phi(k)$ become the RPA pion-creation operator.

If instead Chiral Symmetry was not spontaneously broken in QCD, $M(k)\simeq 0$ and
$\sin \phi(k)\simeq 0$. As a consequence, it is obvious that the chiral
charge would not change the particle content since the second line of
eq.(\ref{chiralcharge}) would vanish, and the first line is made of 
quark and antiquark number operators. Then chiral symmetry would be linearly
realized in Wigner-Weyl mode where hadrons come in degenerate opposite-parity
pairs
\ba \nonumber
[Q_5^a,N^+_i] = \Theta_{ij}^a N^-_j\\ \nonumber 
[Q_5^a,N^-_i] = \Theta_{ij}^a N^+_j \ .
\ea
The parity change follows from the $\sigma\cd \hat{k}$ p-wave present in the
first line of eq.(\ref{chiralcharge}).

In fact, the contemporary realization is that both phenomena are simultaneously
realized in QCD. The vacuum is not annihilated by the chiral charge,
forcing spontaneous symmetry breaking, but the mass gap angle has compact
support and if, in a hadron, the typical quark momentum is high, as illustrated
in figure \ref{fig:restoration}, its wavefunction
is insensitive to Chiral Symmetry Breaking. Therefore one  asymptotically recovers
degenerate Glozman parity doublets.
We will in the following drop the isospin index. 

If a given resonance is high enough in the spectrum so the quarks have a
momentum distribution peaked higher than the support of the gap angle, as in 
figure \ref{fig:restoration}, only the first line of Eq.(\ref{chiralcharge})
is active. $Q_5 \ar N\ra$ contains also three quarks, but one of them
is spin-rotated from $B_{k\lambda}$ to $\sigma\cd\hat{k}_{\lambda
\lambda'}B_{k\lambda'}$. Successive application of the chiral charge spin-rotates
further quarks, changing each time the parity of the total wavefunction.
However the sequence of states ends since 
$\sigma\cd\hat{k}\sigma\cd\hat{k}={\mathbb{I}}$. In fact, starting with
an arbitrary such wavefunction, one generates a quartet
\begin{eqnarray}\nonumber
\ar N_0^P \ra = \sum F^P_{ijk} B^\da_i B^\da_j B^\da_k \ar \Omega \ra 
\\ \nonumber
\ar N_1^{-P} \ra = \frac{1}{3}\sum F^P_{ijk}  \\ \nonumber \left(
\left({\bf \sigma}\cd{\bf \hat{k}}_i B^\da\right)_i B^\da_j B^\da_k +
\ {\rm{permutations}}\right) \ar \Omega \ra \\ \nonumber
\ar N_2^{P} \ra = \frac{1}{3} \sum F^P_{ijk} \\ \nonumber \left(
\left({\bf \sigma}\cd {\bf \hat{k}}_i B^\da\right)_i
\left({\bf \sigma}\cd {\bf \hat{k}}_j B^\da\right)_j 
B^\da_k + \ {\rm{permutations}}
\right) \ar \Omega \ra \\ \nonumber
\ar N_3^{-P} \ra = \sum F^P_{ijk} \\ \nonumber
\left({\bf \sigma}\cd
{\bf \hat{k}}_i B^\da\right)_i
\left({\bf \sigma}\cd {\bf \hat{k}}_j B^\da\right)_j 
\left({\bf \sigma}\cd {\bf \hat{k}}_k B^\da\right)_k \ar \Omega \ra
\end{eqnarray}
that is the natural basis to discuss chiral symmetry restoration in baryons,
through wavefunctions that are linear combinations
$\ar N\ra = \sum c_i\ar N_i \ra$.

Because the Hamiltonian and the chiral charge commute, they can
be diagonalized simultaneously. 

The quartet then separates into two doublets connected by the chiral charge
\ba
Q_5 (N_0-N_2)&=& N_1-N_3   \\ \nonumber
Q_5(N_1 - N_3) &=& N_0-N_2 \\  \nonumber
Q_5 (N_0+3N_2)&=& 3(3N_1+N_3)\\ \nonumber
Q_5(3N_1+N_3) &=& 3(N_0+3N_2)
\ea
Since the quartet can be divided into two two-dimensional irreducible
representations of the chiral group, (with different eigenvalues of
$Q_5^2$, 1 and 9 respectively),
the masses of the two doublets may also be different, and the
interdoublet splitting becomes a dynamical question. 
However, the splitting within the doublet 
\emph{must vanish} asymptotically. This is a prediction following from 
first principles-understanding of QCD alone. Should it not be borne experimentally,
it would falsify the theory.

Of course, parity doubling is a property of a more general class of theories
than QCD. Even for fixed (not running) quark mass, when the typical 
momenta are high enough $\la \ksl \ra >> m$ in the kinetic energy, the
effects of the quark mass are negligible. Parity doubling then
 comes down  to whether
the interaction terms are also chiral symmetry violating or not.

To round off this work, let us look ahead to what the highly excited 
spin spectrum may reveal.
The $J$-dependence of the fall-off of the splittings $M_+-M_-$
is an observable that reveals the underlying chiral theory. If precise
data becomes available at ELSA or Jefferson Lab (note the EBAC, Excited
Baryon Analysis Center effort\cite{JuliaDiaz:2007fa}), in particular 
for the $\Delta_{J}$
with $J=7/2,9/2,11/2$ parity doublets, one should be able to distinguish
between the typical  $1/\sqrt{l}$ fall-off for non-chiral models and the 
faster drop for chiral theories. (Higher yet in the 
spectrum, also the chiral theory may take on the $1/\sqrt{l}$ behavior 
due to the small remaining current quark mass that falls only 
logarithmically)\footnote{
Other authors have argued that flattening of the potential in a 
non-relativistic quark model for large distances due to screening 
(string-breaking) also leads to parity degeneracy \cite{Segovia:2008zz}.
We are preparing an additional paper that 
will provide the necessary detail for chiral models to distinguish 
them.}.

Since the two doublets are closely degenerate, both positive and negative parity
ground states will have a nearby resonance with identical quantum numbers.
Given the width of those states, it is likely they will only be distinguished
by very careful exclusive decay analysis. Meanwhile, if interpreted as
only one resonance, their decay pattern will defy intuition.

It is also worth remarking that the spin-orbit interaction is very small
in the low-lying spectrum, due to cancellations between scalar and vector
potentials and the Thomas precession \cite{isgur}. However, higher in
the spectrum, the vector $\gamma_0\gamma_0$ potential comes forward,
and it is known to present larger spin-orbit splittings than found to
date. Therefore not all splittings in a given baryon shell will disappear
alike: while parity splittings must decrease fast by chiral symmetry,
other spin-orbit splittings will stay constant or even grow. This is 
demanded by a necessary cancellation between $L\cd S$, centrifugal 
forces $l(l+1)$ and tensor forces. This has been explicitly shown for 
mesons in \cite{Bicudo:2007wt}.

It has also been pointed out \cite{Glozman:2005tq,Glozman:2007jt,Nefediev:2008dv}
that the pion decouples from the very excited 
resonances due to the falling overlap between the 
$\Delta*$ wavefunctions and $\sin\phi(k)$ (the pion 
wavefunction in the chiral limit). This might already be
observable in the known widths for pion decays, that decrease even
with larger phase space see table\ref{tablita2}. 
There are lattice calculations addressing 
low-excited baryons \cite{Takahashi:2008fy}, but it 
is still a long way to go until highly excited 
states can be examined.

\begin{table} 
\caption{Total width, exclusive pion-nucleon width and
semiinclusive pion width (decay to one pion plus any other particles
excluding pions) for the ground state $\Delta_J$ resonances. 
All units $MeV$. Data adapted from PDG\cite{Yao:2006px}
\label{tablita2}.}
\begin{tabular}{c|ccc}
$J^P$           & $\Gamma$ & $\Gamma_{\pi N}$ & $\Gamma_{\pi X}$ \\ \hline
$\frac{3}{2}^+$ & 118(2)   & 118(2)           & 118(2)           \\
$\frac{3}{2}^-$ & 300(100) & 50(30)           & 190(90)          \\
$\frac{5}{2}^+$ & 330(60)  & 42(18)           & $<80(20)$           \\
$\frac{5}{2}^-$ & 350(150) & 40(30)           & -           \\
$\frac{7}{2}^+$ & 285(50)  & 115(35)          & 170(30)     \\
$\frac{7}{2}^-$ & 400(150) & 30(20)           & -           \\
$\frac{9}{2}^+$ & 400(150) & 30(20)           & -           \\
$\frac{9}{2}^-$ & 400(180) & 35(25)           & -           \\
$\frac{11}{2}^+$& 450(150) & 50(40)           & -           \\
$\frac{11}{2}^-$&  -       & -                & -           \\
$\frac{13}{2}^+$&  -       & -                & -           \\
$\frac{13}{2}^-$& 400(200) & 20(12)           & -           \\
$\frac{13}{2}^+$& 550(300) & 30(25)           & -           \\
\end{tabular}
\end{table}

{\emph{We are indebted to many colleagues for useful conversations,
among them Leonid Glozman, Emilio Ribeiro, Alexey Nefediev, Atsushi
Hosaka, and Makoto Oka. Work supported by spanish grants
 MCYT FPA 2004-02602,  2005-02327, and Accion Integrada 
Spain-Portugal HP2006-0018. TVC acknowledges the support from the
Fund for Scientific Research - Flanders.}}

\newpage

\end{document}